\documentclass[a4paper,11pt]{article}

\usepackage{pos}
\usepackage{lipsum} 

\title{Towards a more robust reconstruction method for IceCube's real-time program}

\ShortTitle{Towards a more robust reconstruction method for IceCube's real-time program}

\author{The IceCube Collaboration \\{\normalsize \normalfont(a complete list of authors can be found at the end of the proceedings)}\\}

\emailAdd{gsommani@icecube.wisc.edu}
\emailAdd{cristina.lagunas@icecube.wisc.edu}
\emailAdd{hans.niederhausen@icecube.wisc.edu}

\abstract{

Sources of astrophysical neutrinos can potentially be discovered through the detection of neutrinos in coincidence with electromagnetic counterparts. Real-time alerts generated by IceCube play an important role in this search, acting as triggers for follow-up observations with instruments sensitive to electromagnetic signals in various wavelengths. In previous studies, we investigated the treatment of the systematic uncertainties on the reconstruction method currently used in IceCube's real-time program, concluding that a new approach, more robust against systematic variations, is needed. Here we present the state-of-the-art of these analyses, and discuss a modification to an already-existing and reliable reconstruction method that results in an improved solution under many metrics. The proposed reconstruction method is faster, more precise, and significantly less influenced by systematic uncertainties, than the current one. This system provides a more robust estimate of angular uncertainties than the previous algorithm, making it a solid benchmark for real-time event analyses.

\vspace{4mm}
{\bfseries Corresponding authors:}
Giacomo Sommani$^{1*}$, Cristina Lagunas Gualda$^{2}$, Hans Niederhausen$^{3}$\\
{$^{1}$ \itshape Ruhr-Universität Bochum}\\
{$^{2}$ \itshape DESY Zeuthen}\\
{$^{3}$ \itshape  Michigan State University}\\[4mm]
$^*$ Presenter

\ConferenceLogo{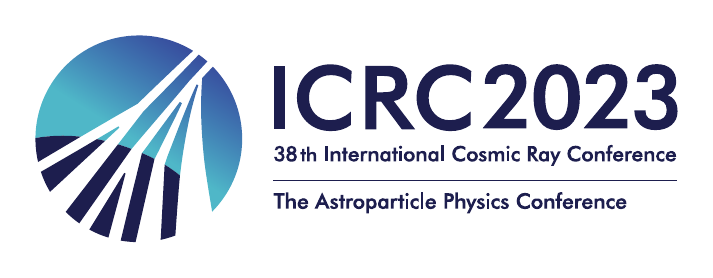}

\FullConference{The 38th International Cosmic Ray Conference (ICRC2023)\\ 26 July -- 3 August, 2023\\ Nagoya, Japan}
}

\begin{document}

\maketitle

\section{Introduction}\label{sec:intro}

IceCube is a cubic-kilometer scale neutrino detector located at the South Pole~\cite{Aartsen:2016nxy}. It detects the Cherenkov photons induced by the charged particles resulting from the interaction of neutrinos with the Antarctic ice. IceCube discovered an isotropic flux of astrophysical neutrinos in 2013~\cite{diffuse2013}, but the sources of those neutrinos are still largely unknown. One of the most promising possibilities is that high-energy neutrinos are produced in very powerful, transient phenomena. To find such source candidates, a prompt search for an electromagnetic counterpart after the neutrino detection is required. The IceCube Real-time System selects high-energy neutrinos with a high probability of being astrophysical in origin. For these neutrinos an alert is immediately sent out to other observatories to search for electromagnetic counterparts. It was established in 2016~\cite{Aartsen:2016lmt}, and the selection criteria were revised in 2019~\cite{Blaufuss:2019fgv}. This system lead to the first evidence for neutrino emission from an extragalactic object, the blazar TXS 0506+056, which was followed up after the initial report of a 290 TeV energy track-like neutrino event~\cite{doi:10.1126/science.aat2890}. The work presented here focuses on the through-going track selection of the Real-time System, which comprises muon neutrinos that undergo charged-current interactions and produce muons.

The direction of each neutrino alert event is reconstructed first with the SplineMPE~\cite{abbasi2021muontrack} method, and the information is distributed in a GCN Notice\footnote{\url{https://gcn.nasa.gov/notices}}. After the automated GCN Notice is sent out, a more sophisticated and computationally expensive reconstruction method based on and referred to as Millipede~\cite{energyreco} starts. This method provides a likelihood landscape on a Hierarchical Equal Area isoLatitude Pixelization (HEALPix)~\cite{Gorski_2005} grid, where the reconstructed direction is given as the pixel with the maximum likelihood. From the scan one can derive error contours at different confidence levels (CLs) as well. To account for systematic uncertainties, the 50\% and 90\% error contours calculated with Wilks' theorem are scaled up with a fixed set of values, determined by simulations \cite{panstarrs}. The best-fit position and the minimum rectangle that encapsulates the 90\% CL error contour are distributed in a GCN Circular\footnote{\url{https://gcn.nasa.gov/circulars}}, a few hours after the initial alert.

In a previous study \cite{gualda2021studies}, we checked the validity of said scaling values, and concluded that the error contours calculated with them do not have always the expected coverage. For that, a set of high-energy neutrinos were simulated, varying the ice model parameters during the simulation within a range based on calibration data. Here, we present an alternative solution to the problem. It utilizes a similar simulation scheme but relies on the fast angular reconstruction algorithm called SplineMPE for the reconstruction of the events, which allows high-statistics studies of systematic uncertainties not possible before because of the high computational cost of Millipede.

In Section~\ref{sec:scansplinempe}, we explain the new implementation of SplineMPE used for the results presented in this proceeding. In Section~\ref{sec:gms}, we apply the new scan to the set of simulated neutrinos from \cite{gualda2021studies} for comparison. Section~\ref{sec:benchmarksim} and Section~\ref{sec:benchmarkresults} show the new simulation data set and the corresponding results. Lastly, in Section~\ref{sec:conclusions} we discuss the future plans and conclude. 

\section{Implementation of the likelihood scan for SplineMPE}\label{sec:scansplinempe}

IceCube estimates the 50\% and 90\% CL error contours for the real-time alerts using the reconstruction algorithm Millipede, as mentioned in Section~\ref{sec:intro} (explained in detail in \cite{gualda2021studies}). Nevertheless, most analyses searching for neutrino sources use another angular reconstruction method, called SplineMPE~\cite{abbasi2021muontrack}. The main difference between the two algorithms consists on the type of muon energy losses that they consider, and therefore, on the light emission. SplineMPE assumes a continuous Cherenkov light deposition, while Millipede expects stochastic energy losses to happen along the muon trajectory. The latter case is a more realistic description of what is really happening in the detector. However, for a directional reconstruction, the assumption of continuous light emission of the SplineMPE algorithm could already be sufficient. This scenario would significantly simplify the fitting process and reduce the dependencies on the systematic uncertainties of the detector (most importantly the south-pole ice models). 

Both algorithms converge easily into local minima during the likelihood minimization process. To ensure convergence to the true minimum, the Millipede reconstruction method is performed in a likelihood sky scan~\cite{gualda2021studies}, where the sky is divided in pixels of equal area with HEALPix~\cite{Gorski_2005}. For each pixel, a track hypothesis is assumed, and the likelihood that this track resembles the data is calculated by fitting the observed energy losses to those predicted with Millipede. The pixel with the maximum likelihood then gives the reconstructed direction. A similar likelihood scan was developed for SplineMPE, also based on a HEALPix grid. Millipede's likelihood depends on the track vertex, the direction, and the energy. Therefore, for each pixel the negative logarithm of the likelihood is minimized over vertex and energy. SplineMPE's likelihood does not depend on the energy, hence the minimization is only with respect to the vertex.

The likelihood scan can be used to derive the uncertainty contours with different confidence levels. The method is similar to the process explained in~\cite{gualda2021studies}. First, events that are similar to one another are simulated. Then, they are reconstructed with SplineMPE, and the difference in log-likelihood ($\Delta$LLH) between the simulated direction and the reconstructed direction is calculated for all events in the data set. The distribution of $\Delta$LLH is computed, and the 50\% and 90\% containment values are used to derive the 50\% and 90\% error contours of a scan by searching for the pixels that have the same difference in log-likelihood to the best-fit position. 

\section{Results on previous simulations}\label{sec:gms}

\begin{figure}
  \centering
  \includegraphics[width=0.38\linewidth]{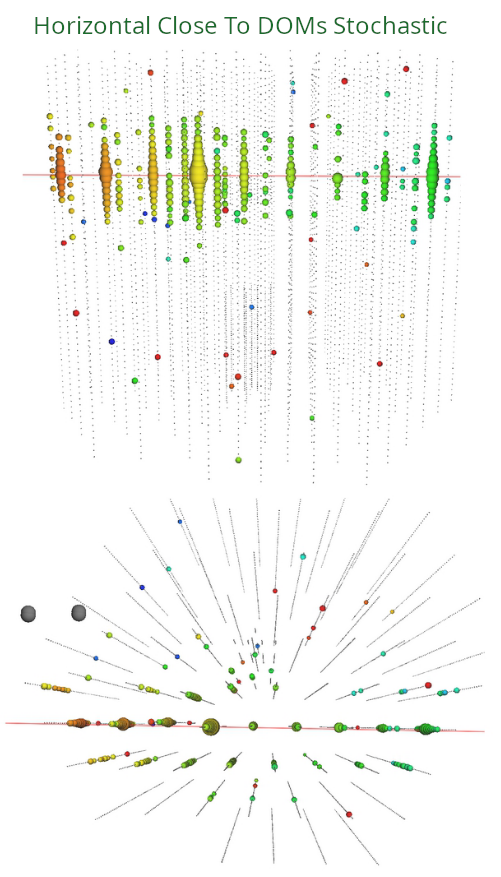}
  \includegraphics[width=0.38\linewidth]{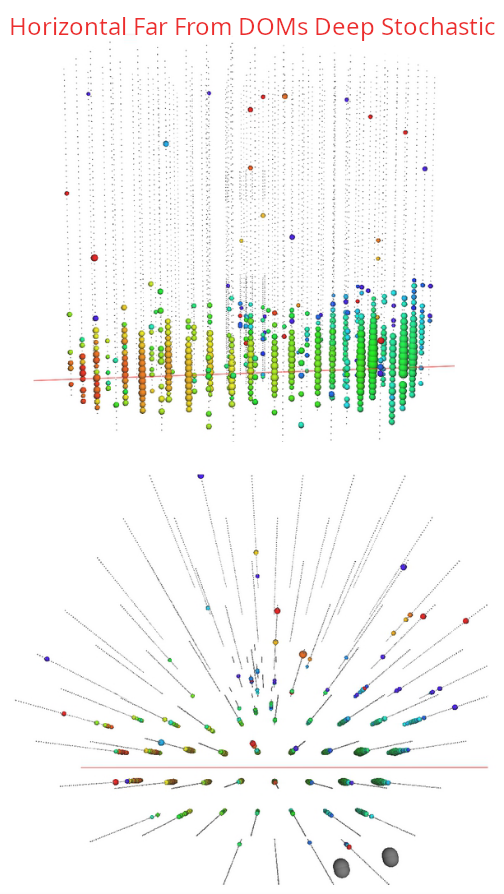}
  \caption{Left: An example event from the \textit{Shallow, Close To DOMs} category. Right: An example event from the \textit{Deep, Far From DOMs} category. Each colored sphere represents a DOM that has detected Cherenkov light from the event, whose radius is proportional to the deposited charge. The colors show the temporal evolution of the track, from early photons (red) to late photons (blue). These are simulated events.}
  \label{fig:example-eventviews}
\end{figure}

Here we apply the SplineMPE likelihood scan explained in Section~\ref{sec:scansplinempe} to the simulated data produced for a previous work~\cite{gualda2021studies}. These simulations consist of through-going tracks divided into eleven different categories, all of them with the same neutrino energy, $150\,$TeV, close to the median energy of IceCube's alert events~\cite{gualda2021studies}. For each category, there are 100 simulated events, all assuming the same true neutrino direction. These 100 events have been simulated by fixing the muon energy losses along the track and performing the propagation of the Cherenkov photons varying the parameters of the ice model with the simulation tool \textit{SnowStorm}~\cite{Aartsen_2019}. Each simulation of the same event has a unique set of ice properties. The variation of the ice properties permits to take into account the systematic uncertainties. The categories are divided in the following way, with two main classes:

\begin{enumerate}
    \item \textbf{Horizontal}, tracks with zenith angle $\theta=90^\circ$ and azimuth angle $\phi=9^\circ$ in the detector coordinate system (these events travel parallel to \textit{corridors} between strings). The simulated events in this class are divided according to the following characteristics:
    \begin{enumerate}
        \item \underline{Distance from DOMs}, feature related to the distance of the muon track to the Digital Optical Modules (DOMs)~\cite{Aartsen:2016nxy}, the main detection unit in IceCube. They are \textit{Close To DOMs}, if the track passes close to the individual DOMs, or \textit{Far From DOMs}, if it passes through a \textit{corridor} between strings (see Figure~\ref{fig:example-eventviews}).
        
        \item \underline{Depth}, feature related to the depth of the muon track in the glacier. \textit{Shallow} tracks pass through the upper part of the detector, above the dust layer\footnote{the ice at depths between 1970 and 2100 m has a relatively short absorption length, and is known as the “dust layer”.}. \textit{Deep} tracks pass through the lower part of the detector, below the dust layer.
    
        \item \underline{Stochasticity}, feature related to the light emission. If the light emission is uniform along the muon track, the category is \textit{Smooth}. If there is at least one big stochastic energy loss along the muon track, then it is \textit{Stochastic}.
    \end{enumerate}
    
    \item \textbf{Upgoing}, with zenith angle $\theta=130^\circ$ and azimuth angle $\phi = 0^\circ$. The simulated events in this class are divided in three different categories: \textit{Smooth}, \textit{Stochastic} (with the same definition as for the \textit{Horizontal} class) and \textit{Repeated Muon Propagation}. The latter is different from all the other categories, since the muon propagation is repeated for each one of the 100 simulations, leading to different energy loss patterns.
    
\end{enumerate}

\begin{table}
\centering
\begin{tabular}{ l | c c c c | c }

    & \multicolumn{2}{c}{\textbf{$\Delta$LLH}} & \multicolumn{2}{c}{\textbf{Ang. Dist. [deg]}} & \textbf{KS test on $\Delta$LLH}\\

    \textbf{Category} & \textbf{50\%} & \textbf{90\%} & \textbf{50\%} & \textbf{90\%} & \textbf{and $\chi^2$ distr.}\\
    \hline
    \hline
    \textit{Up. Rep. Muon Prop.} & 2.1 & 8.7 & 0.07 & 0.13 & $3.5\times10^{-4}$\\
    \textit{Up. Smooth} & 1.5 & 5.1 & 0.06 & 0.13 & $3.9\times10^{-5}$\\
    \textit{Up. Stoch.} & 2.8 & 8.2 & 0.07 & 0.14 & $7.6\times10^{-8}$\\
    \hline
    \textit{Hor. CloseToDOMs Shall. Smooth} & 2.0 & 5.8 & 0.07 & 0.13 & $1.1\times10^{-7}$\\
    \textit{Hor. CloseToDOMs Shall. Stoch.} & 1.4 & 4.8 & 0.07 & 0.14 & $4.8\times10^{-1}$\\
    \textit{Hor. CloseToDOMs Deep Smooth} & 2.2 & 6.7 & 0.08 & 0.13 & $8.6\times10^{-4}$\\
    \textit{Hor. CloseToDOMs Deep Stoch.} & 2.9 & 9.0 & 0.08 & 0.17 & $1.1\times10^{-7}$\\
    \hline
    \textit{Hor. FarFromDOMs Shall. Smooth} & 0.8 & 2.9 & 0.19 & 0.40 & $4.0\times10^{-5}$\\
    \textit{Hor. FarFromDOMs Shall. Stoch.} & 2.1 & 5.7 & 0.32 & 0.59 & $1.2\times10^{-2}$\\
    \textit{Hor. FarFromDOMs Deep Smooth} & 9.3 & 25.8 & 0.53 & 0.90 & $\ll10^{-4}$\\
    \textit{Hor. FarFromDOMs Deep Stoch.} & 2.9 & 11.0 & 0.34 & 0.68 & $2.2\times10^{-9}$\\
    \hline
    Wilks' Theorem & 1.4 & 4.6 & - & - & - \\
    \hline
    $3\sigma$ KS test & - & - & - & - & $1.4\times10^{-3}$ \\
    $3\sigma$ KS test trial-corrected & - & - & - & - & $1.2\times10^{-4}$ 
\end{tabular}
\caption{Results of the analysis on the simulations performed using SplineMPE. Each category contains 100 reconstructed events. The last column shows the results of the Kolmogorov-Smirnov tests comparing the $\Delta$LLHs of the different simulation categories with the $\chi^2$~distribution. The table shows also the $\Delta$LLH levels assuming Wilks' theorem and the $3\sigma$ values (pre-trial and post-trial) for the Kolmogorov-Smirnov test.}
\label{tab:cvs_gms}
\end{table}

\begin{figure}
  \centering
  \includegraphics[width=0.8\linewidth]{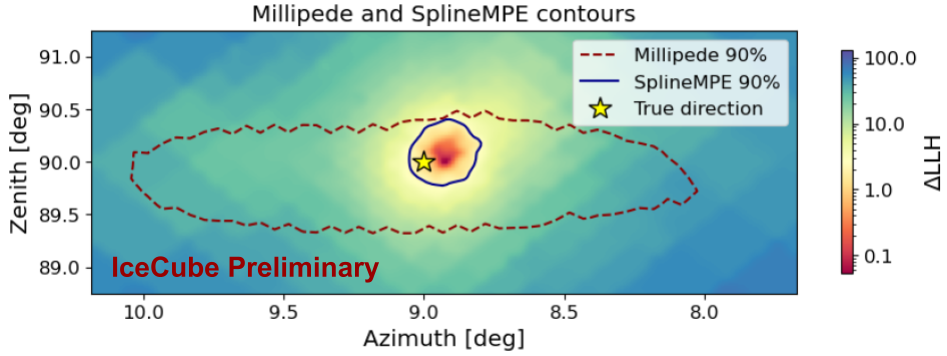}
    \caption{Comparison between the contours with 90\% coverage computed for one simulation using the two algorithms Millipede and SplineMPE. The colors in the background show the values of $\Delta$LLH for each pixel of the SplineMPE likelihood scan. A Gaussian interpolation between the pixels was applied to smooth the picture and the contours. The category of the simulated event is \textit{Horizontal CloseToDOMs Shallow Smooth}.}
    \label{fig:mill_vs_spl}
\end{figure}

Table~\ref{tab:cvs_gms} shows the results of the scans for all simulation sets with SplineMPE. In addition to log-likelihood ratio ($\Delta$LLH) levels, the table shows the 50\% and 90\% containment of the angular distances between the reconstructed direction and the simulated direction. The $\Delta$LLH values are much smaller than the ones obtained with Millipede (see~\cite{gualda2021studies}). Moreover, they are closer to the values that one would obtain with Wilks' theorem, i.e., the 50\% and 90\% containment values of a $\chi^2$ distribution with two degrees of freedom. With only 100 events determining the results, the $\Delta$LLH levels are similar across most of the various categories of the simulations. In Figure~\ref{fig:mill_vs_spl}, an example SplineMPE scan result of a simulated event is shown. The dashed and solid lines represent the 90\% CL error contours computed using the scaling values calculated with Millipede and SplineMPE, respectively. The Millipede contour is obtained from a separate scan, not shown in the figure. 

A Kolmogorov-Smirnov (KS) test can be used to verify whether the $\Delta$LLH for the various categories follow the $\chi^2$~distribution with two degrees of freedom expected from Wilks' theorem if there are no systematic uncertainties. The results, reported in Table~\ref{tab:cvs_gms}, exclude the $\Delta$LLH following a $\chi^2$~distribution for six categories.

\section{The real-time benchmark simulations}\label{sec:benchmarksim}

The simulations introduced above provided the first step in the study of the impact of systematic uncertainties on the angular reconstruction method employed in the Real-time System. However, the simulated tracks are brighter, longer and more uniform than those of most alert events. With the goal of moving towards a more realistic sample, we use a previously simulated data set of neutrino alert events, more heterogeneous in energies and directions, which we call \textit{the real-time benchmark simulations} in the following. Originally, these were created to calculate the probability of each individual neutrino to be of astrophysical origin~\cite{Aartsen:2016lmt, Blaufuss:2019fgv}. Here, we randomly select 100 neutrino events from this simulated dataset, that each would have generated an alert in the real-time system, had those been observed in real-data. We then re-simulated each individual event 100 times following the method used for the real-time benchmark simulations is the same as for the previous simulations (described in \cite{gualda2021studies} and summarized in Section~\ref{sec:gms}). The muon propagation is fixed for each simulation of the same event, i.e. the energy losses and simulated direction are kept the same, and only the photon propagation is repeated with varying ice model parameters using \textit{SnowStorm}, for example enhancing scattering by 10\%.

\section{Results on the real-time benchmark simulations}\label{sec:benchmarkresults}

\begin{figure}
    \centering
    \includegraphics[width=0.7\linewidth]{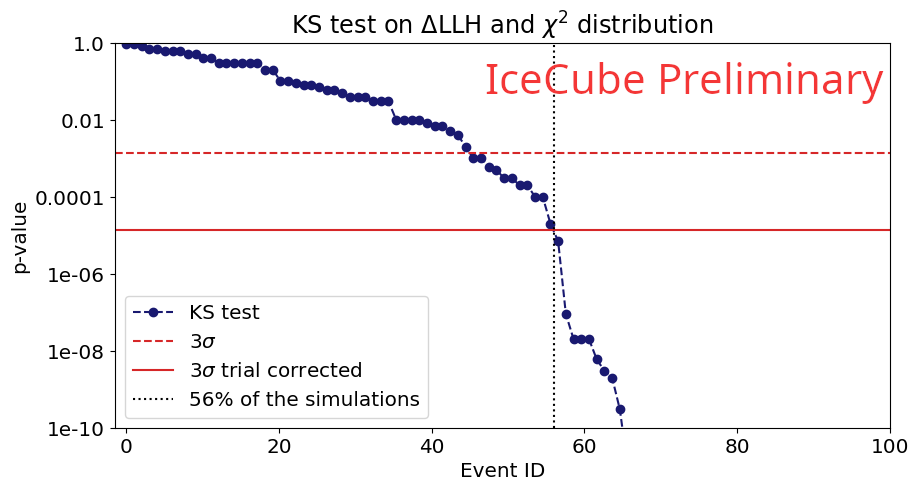}
    \caption{Results of the Kolmogorov-Smirnov tests comparing the $\Delta$LLH distribution of the real-time benchmark simulations and the $\chi^2$ distribution. The Event ID is chosen to order the simulated events according to the result of the Kolmogorov-Smirnov test. The trial correction takes into account the number of trials. The number of simulated events is 100, if the p-value is smaller than 10$^{-10}$ the result does not appear in the plot.}
    \label{fig:ks-benchmark}
\end{figure}

\begin{figure}
    \centering
    \includegraphics[width=0.75\linewidth]{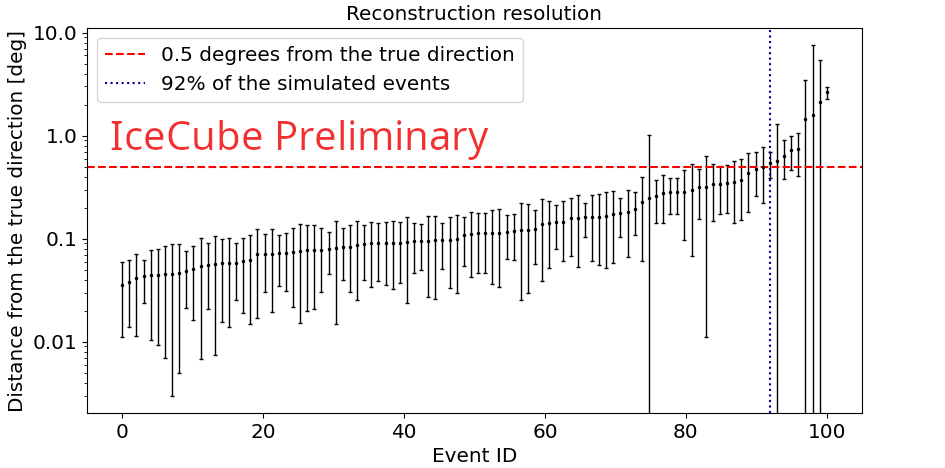}
    \caption{Angular distance between the reconstructed and true direction for the various real-time benchmark simulations. The Event ID is assigned ordering the simulated events on the angular distance from the true direction. Each simulation is resimulated 100 times. In the plot are indicated the average distances and the standard deviations.}
    \label{fig:resolution-benchmark}
\end{figure}

The real-time benchmark simulations were reconstructed in the same way as the previous simulations using SplineMPE. For each of the 100 events the $\Delta$LLH levels of their resimulations were calculated. Also in this case, these levels are compared with the $\chi^2$ distribution using a KS test, see Figure~\ref{fig:ks-benchmark}. For $\sim50\%$ of the events there is a good agreement between the $\Delta$LLH levels of the resimulations and the expected $\chi^2$~distribution. However, for the other half of the events the agreement is poor hinting at the influence of of deficiencies in our likelihood model. Figure~\ref{fig:resolution-benchmark} shows the average distances between the reconstructed and the true directions for the resimulations of each of the 100 events.

For many events, even if the $\Delta$LLH distributions are not compatible with Wilks' theorem (see Figure~\ref{fig:ks-benchmark}), the direction is reconstructed relatively close to the expected position. About $90\%$ of the events have an average distance from the true direction that is less than 0.5$^\circ$. Moreover, about $70\%$ of the events have an average distance from the true direction that is less than $0.2^\circ$. We also simulated the same events but without varying the ice model parameters with \textit{SnowStorm} and we obtained the same results. We note that the reconstruction algorithm SplineMPE performs a likelihood minimization using an older version of the ice model~\cite{Aartsen_2013}, while the simulations were produced using a more recent ice model~\cite{chirkin2019light}, further illustrating the robustness of SplineMPE against ice systematics. The good performances in resolution of SplineMPE, and the agreement with the expected distribution of the $\Delta$LLH for at least 50\% of the events makes this method a leading candidate to replace the current method.

\section{Conclusion}\label{sec:conclusions}

The SplineMPE algorithm implemented with a likelihood scan returns promising results. Even if in some cases they are still diverging, the $\Delta$LLH levels are in much better agreement with the $\chi^2$ distribution predicted by Wilks' theorem than the ones obtained using the current method. Moreover, the reconstructed directions are close to the true ones ($\sim90\%$ within $0.5^\circ$). All these reconstructions were performed using a different ice model to the one used for the simulation of the events. This illustrates the robustness of the algorithm against systematic uncertainties. All these factors go together with a reconstruction performed by SplineMPE that is simpler and faster, with an estimate of $\sim3$ CPU/hrs per scan, compared to the $10^4$ CPU/hrs required by Millipede. 
In future, we plan to study the possibility of a fully simulation derived uncertainty contour of the single alert in real-time \cite{Jannis:2023icrc}. In this case, it would, for example, be possible to calibrate the $\Delta$LLH levels in real-time, to produce contours with the correct coverage. Further studies will also be performed to improve the SplineMPE algorithm to have a model that produces $\Delta$LLH distributions compatible with Wilks' theorem in all cases.

\bibliographystyle{ICRC}

\setlength{\bibsep}{0.8ex}
\bibliography{references}

\clearpage

\section*{Full Author List: IceCube Collaboration}

\scriptsize
\noindent
R. Abbasi$^{17}$,
M. Ackermann$^{63}$,
J. Adams$^{18}$,
S. K. Agarwalla$^{40,\: 64}$,
J. A. Aguilar$^{12}$,
M. Ahlers$^{22}$,
J.M. Alameddine$^{23}$,
N. M. Amin$^{44}$,
K. Andeen$^{42}$,
G. Anton$^{26}$
% ...
\\
\\
% ...
$^{12}$ Universit{\'e} Libre de Bruxelles, Science Faculty CP230, B-1050 Brussels, Belgium \\
% ...
$^{17}$ Department of Physics, Loyola University Chicago, Chicago, IL 60660, USA \\
$^{18}$ Dept. of Physics and Astronomy, University of Canterbury, Private Bag 4800, Christchurch, New Zealand \\
% ...
$^{22}$ Niels Bohr Institute, University of Copenhagen, DK-2100 Copenhagen, Denmark \\
$^{23}$ Dept. of Physics, TU Dortmund University, D-44221 Dortmund, Germany \\
% ...
$^{26}$ Erlangen Centre for Astroparticle Physics, Friedrich-Alexander-Universit{\"a}t Erlangen-N{\"u}rnberg, D-91058 Erlangen, Germany \\
% ...
$^{40}$ Dept. of Physics and Wisconsin IceCube Particle Astrophysics Center, University of Wisconsin{\textendash}Madison, Madison, WI 53706, USA \\
% ...
$^{42}$ Department of Physics, Marquette University, Milwaukee, WI, 53201, USA \\
% ...
$^{44}$ Bartol Research Institute and Dept. of Physics and Astronomy, University of Delaware, Newark, DE 19716, USA \\
% ...
$^{63}$ Deutsches Elektronen-Synchrotron DESY, Platanenallee 6, 15738 Zeuthen, Germany  \\
$^{64}$ Institute of Physics, Sachivalaya Marg, Sainik School Post, Bhubaneswar 751005, India \\
% ...

\subsection*{Acknowledgements}

\noindent
The authors gratefully acknowledge the support from the following agencies and institutions:
USA {\textendash} U.S. National Science Foundation-Office of Polar Programs,
U.S. National Science Foundation-Physics Division,
U.S. National Science Foundation-EPSCoR,
Wisconsin Alumni Research Foundation,
Center for High Throughput Computing (CHTC) at the University of Wisconsin{\textendash}Madison,
Open Science Grid (OSG),
Advanced Cyberinfrastructure Coordination Ecosystem: Services {\&} Support (ACCESS),
Frontera computing project at the Texas Advanced Computing Center,
U.S. Department of Energy-National Energy Research Scientific Computing Center,
Particle astrophysics research computing center at the University of Maryland,
Institute for Cyber-Enabled Research at Michigan State University,
and Astroparticle physics computational facility at Marquette University;
Belgium {\textendash} Funds for Scientific Research (FRS-FNRS and FWO),
FWO Odysseus and Big Science programmes,
and Belgian Federal Science Policy Office (Belspo);
Germany {\textendash} Bundesministerium f{\"u}r Bildung und Forschung (BMBF),
Deutsche Forschungsgemeinschaft (DFG),
Helmholtz Alliance for Astroparticle Physics (HAP),
Initiative and Networking Fund of the Helmholtz Association,
Deutsches Elektronen Synchrotron (DESY),
and High Performance Computing cluster of the RWTH Aachen;
Sweden {\textendash} Swedish Research Council,
Swedish Polar Research Secretariat,
Swedish National Infrastructure for Computing (SNIC),
and Knut and Alice Wallenberg Foundation;
European Union {\textendash} EGI Advanced Computing for research;
Australia {\textendash} Australian Research Council;
Canada {\textendash} Natural Sciences and Engineering Research Council of Canada,
Calcul Qu{\'e}bec, Compute Ontario, Canada Foundation for Innovation, WestGrid, and Compute Canada;
Denmark {\textendash} Villum Fonden, Carlsberg Foundation, and European Commission;
New Zealand {\textendash} Marsden Fund;
Japan {\textendash} Japan Society for Promotion of Science (JSPS)
and Institute for Global Prominent Research (IGPR) of Chiba University;
Korea {\textendash} National Research Foundation of Korea (NRF);
Switzerland {\textendash} Swiss National Science Foundation (SNSF);
United Kingdom {\textendash} Department of Physics, University of Oxford.

\end{document}